\documentclass{article}
\usepackage[preprint]{spconf,amsmath,graphicx, amssymb}
\usepackage{url}
\usepackage{pgfplots}
\usepackage{tikz}
\usepackage{tkz-fct}
\usepackage{tkz-euclide}
\usepackage[hidelinks]{hyperref}

\copyrightnotice{\fbox{\parbox{\dimexpr\textwidth-\fboxsep-\fboxrule\relax}{\footnotesize \textcopyright 2021 IEEE. Personal use of this material is permitted.
					Permission from IEEE must be obtained for all other uses, in any current or future
					media, including reprinting/republishing this material for advertising or promotional
					purposes, creating new collective works, for resale or redistribution to servers or
					lists, or reuse of any copyrighted component of this work in other works.
					DOI: \href{https://doi.org/10.1109/ICASSP39728.2021.9413684}{10.1109/ICASSP39728.2021.9413684}
}}}

\newlength\fwidth
\tikzset{cross/.style={cross out, draw=black, minimum size=2*(#1-\pgflinewidth), inner sep=0pt, outer sep=0pt},cross/.default={4pt}}

\makeatletter
\g@addto@macro\normalsize{\fontsize{9.5pt}{11.38pt} \selectfont}
\makeatother

\title{Frame Rate Up-Conversion using Key Point Agnostic Frequency-Selective Mesh-to-Grid Resampling}
\name{Viktoria Heimann, Andreas Spruck, and Andr\'e Kaup }
\address{ 
	Multimedia Communications and Signal Processing\\
 	Friedrich-Alexander Universit\"at, Erlangen-N\"urnberg 
}

\begin{document}
%
\maketitle
\begin{abstract}
High frame rates are desired in many fields of application. As in many cases the frame repetition rate of an already captured video has to be increased, frame rate up-conversion (FRUC) is of high interest. We conduct a motion compensated approach. From two neighboring frames, the motion is estimated and the neighboring pixels are shifted along the motion vector into the frame to be reconstructed. For displaying, these irregularly distributed mesh pixels have to be resampled onto regularly spaced grid positions. We use the model-based key point agnostic frequency-selective mesh-to-grid resampling (AFSMR) for this task and show that AFSMR works best for applications that contain irregular meshes with varying densities. AFSMR gains up to 3.2~dB in contrast to the already high performing frequency-selective mesh-to-grid resampling (FSMR). Additionally, AFSMR increases the run time by a factor of 11 relative to FSMR.
\end{abstract}
\section{Introduction}
\label{sec:intro}
The generation of additonal video frames is of high importance in many applications. In entertainment industry, the demand for additional video frames is high, e.g., for slow motion generation, for novel view synthesis, or in frame recovery in video streaming. Additionally, varying repetition rates of captured videos and replaying devices ask for frame rate up-conversion (FRUC) \cite{Ohm_2016}. Furthermore, high frame rates are also desired in video surveillance and automotive applications. Hardware that can fullfill these needs is too expensive in many scenarios so that the additional frames have to be generated artificially. \\
\begin{figure}[th]
\centering
\resizebox{\linewidth}{!}{
\definecolor{mycolor1}{rgb}{0.7, 0.7, 0.7}%

\begin{tikzpicture}
    \tkzInit[xmin=0,xmax=12,ymin=-2,ymax=8] 
    \tkzClip[space=.5] 
    \tikzstyle{arrow} = [thick,->,>=stealth]
    \tkzDefPoint(0,0){A} 
    \tkzDefPoint(0,4){B} 
    \tkzDefPoint(3,2){C} 
    \tkzDefPointWith[colinear= at C](A,B) \tkzGetPoint{D}
    \tkzDefPoint(4.5,0){E} 
    \tkzDefPoint(4.5,4){F} 
    \tkzDefPoint(7.5,2){G} 
    \tkzDefPointWith[colinear= at G](E,F) \tkzGetPoint{H}    
    \tkzDefPoint(9,0){I} 
    \tkzDefPoint(9,4){M} 
    \tkzDefPoint(12,2){K} 
    \tkzDefPointWith[colinear= at K](I,M) \tkzGetPoint{L}
    \tkzDefPoint(0.75, 3.5){startOne}
    \tkzDefPoint(6,3){end}
    \tkzDefPoint(11.25,2.5){startTwo}
    \tkzDrawPolygon[line width = 0.5mm](A,C,D,B)
    \tkzDrawPolygon[line width = 0.5mm, color = mycolor1](E,G,H,F)
    \tkzDrawPolygon[line width = 0.5mm](I,K,L,M)
         \draw [decoration={markings,mark=at position 1 with {\arrow[scale=3,>=stealth]{>}}},postaction={decorate}](startOne) -- node[above] { $ [\Delta m, \Delta n] \in \mathcal{O}^{(c-1\rightarrow c+1)} $} (startTwo); 
	\node at (1, -1.5) { $F^{(c-1)}$};
	\node at (5.5, -1.5) { $F^{(c)}$};
	\node at (10, -1.5) { $F^{(c+1)}$};
	
\end{tikzpicture}
}
\caption{\label{Fig:framework_fruc} The framework for FRUC applications using unidirectional motion compensation. The vector $[\Delta m, \Delta n] \in \mathcal{O}^{(c-1\rightarrow c+1)} $ is a representant of the set of motion vectors in forward direction. The frame $F^{(c)}$ in gray has to be reconstructed. }
\end{figure}
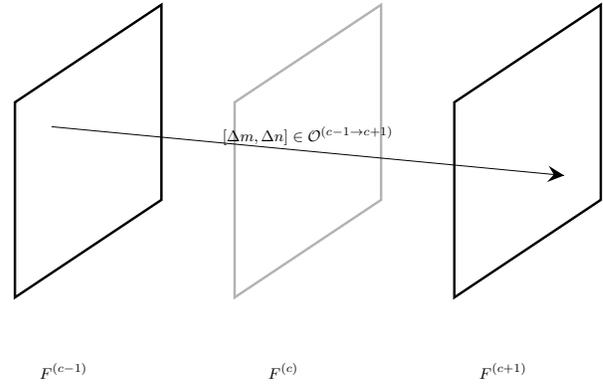
There exist three categories of FRUC approaches: the recently developed approaches using deep learning, non-motion compensated, and motion compensated (MC). The recent approaches using deep learning mostly combine motion compensation and image interpolation into one framework \cite{Bao_2019, Zhang_2020}. Non-motion compensated approaches like the projection of the temporally closest original frame can be used as fallback solutions \cite{Ohm_2016}. Also numerous MC methods are proposed in literature, e.g. MC shifting, MC averaging \cite{Haan_2010}. Additionally, more sophisticated MC approaches are published \cite{Batz_2017, Kaviani_2016}. In these approaches, bidirectional motion estimation is used for motion compensation. On the decoder side of transmission, some scenarios demand for unidirectional motion estimation and consequently, motion compensation. Furthermore, motion field estimators based on neural networks like \cite{Teed_2020} estimate the flow just in one direction as well. Hence, using unidirectional motion compensation is the most general approach of synthesizing additional video frames and increasing the frame rate. As in these cases the number of pixels that can be used for interpolation is only half the number of pixels than for bidirectional motion estimation cases, it is crucial to use the best interpolation technique available. Thus, we propose to use Key Point Agnostic Frequency-Selective Mesh-to-Grid Resampling (AFSMR) to solve the resampling problem. ASFMR already showed high performance for affine transforms \cite{Heimann_2020_MMSP}. Now, we demonstrate the performance of AFSMR for meshes with varying densities as they result from MC. \\
Our framework for FRUC using unidirectional motion compensation is further described in the next section. In Section \ref{sec:interpolation} our AFSMR algorithm is explained in further detail for the application of FRUC. Subsequently, the simulation results are shown and discussed in Section \ref{sec:evaluation}. Section \ref{sec:conclusion} concludes this paper.

\section{Our approach to FRUC}
\label{sec:fruc}
In frame rate up-conversion, the current view in frame $F^{(c)}$ should be generated. We use a motion compensated approach that uses unidirectional motion estimation. Therefore, we take the two neighboring frames $F^{(c-1)}$ and $F^{(c+1)}$ into account. For the previous frame, we define the matrices $\boldsymbol M^{(c-1)}$, $\boldsymbol N^{(c-1)}$, and $\boldsymbol V^{(c-1)}$ containing the horizontal coordinates, vertical coordinates, and color values, respectively. This triplet forms the set of points in $F^{(c-1)}$, thus
\begin{equation}
\label{eq:pointset}
\mathcal{F}^{(c-1)}=\{(\boldsymbol M^{(c-1)}, \boldsymbol N^{(c-1)}, \boldsymbol V^{(c-1)})\}.
\end{equation} 
Analogously, we conduct the same approach for the posterior frame $F^{(c+1)}$, so that the set of points for the posterior frame is given to 
\begin{equation}
\label{eq:pointset_c+1}
\mathcal{F}^{(c+1)}=\{(\boldsymbol M^{(c+1)}, \boldsymbol N^{(c+1)}, \boldsymbol V^{(c+1)})\}.
\end{equation}
In this paper, we use the most general approach for motion estimation and thus, apply unidirectional motion estimation in forward direction. Our framework is visualized in Figure~\ref{Fig:framework_fruc}. Hence, we estimate the motion in forward direction from frame $F^{(c-1)}$ to $F^{(c+1)}$. Therefore, we define the set of motion vectors 
\begin{equation}
\mathcal{O}^{(c-1\rightarrow c+1)} = \{\Delta\boldsymbol M^{(c-1\rightarrow c+1)}, \Delta \boldsymbol N^{(c-1\rightarrow c+1)})\},
\end{equation}
where $\Delta\boldsymbol M^{(c-1\rightarrow c+1)}$ gives the displacement in horizontal direction and $\Delta\boldsymbol N^{(c-1\rightarrow c+1)}$ in vertical direction for every pixel of $\mathcal{F}^{(c-1)}$. To compensate for the forward motion in $F^{(c-1)}$, we get
\begin{align}
\boldsymbol M^{(c-1)}_{\text{MC}} &= \boldsymbol M^{(c-1)} + \frac{1}{2} \Delta\boldsymbol M^{(c-1\rightarrow c+1)},  \text{and} \\
\boldsymbol N^{(c-1)}_{\text{MC}} &= \boldsymbol N^{(c-1)} + \frac{1}{2} \Delta\boldsymbol N^{(c-1\rightarrow c+1)}.
\end{align}
Hence, the motion compensated point set results in 
\begin{equation}
\mathcal{F}^{(c-1)}_{\text{MC}}=\{(\boldsymbol M^{(c-1)}_{\text{MC}}, \boldsymbol N^{(c-1)}_{\text{MC}}, \boldsymbol V^{(c-1)})\}.
\end{equation}
This point set ist then equivalent to the point set in the current view $F^{(c)}$, thus
\begin{equation}
\mathcal{F}^{(c)} = \mathcal{F}^{(c-1)}_{\text{MC}}.
\end{equation}
Due to the motion compensation, these points are distributed arbitrarily with varying density on non-integer positions within the frame. We will refer to these points as mesh in the following. The problem arises that the mesh points cannot be displayed on a digital screen nor can they be stored efficiently. Hence, a resampling process has to be conducted that resamples the mesh points onto regular integer positions which we refer to as grid positions in the following.

\section{Interpolation}
\label{sec:interpolation}
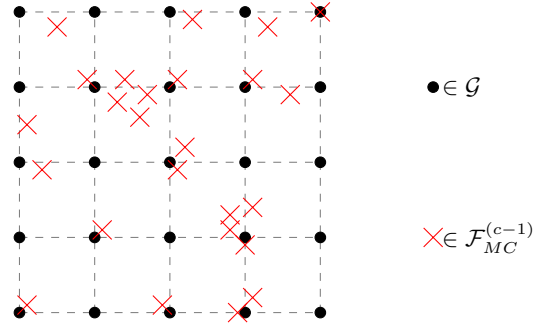
\begin{figure}
\centering
\begin{tikzpicture}
	\tkzInit[xmax=4, ymax=4]
	\begin{scope}[dashed]
		\tkzGrid
	\end{scope}
	\draw[fill=black](0,0)circle(2pt);
	\draw[fill=black](0,1)circle(2pt);
	\draw[fill=black](0,2)circle(2pt);
	\draw[fill=black](0,3)circle(2pt);	
	\draw[fill=black](0,4)circle(2pt);	
	\draw[fill=black](1,0)circle(2pt);
	\draw[fill=black](1,1)circle(2pt);	
	\draw[fill=black](1,2)circle(2pt);
	\draw[fill=black](1,3)circle(2pt);	
	\draw[fill=black](1,4)circle(2pt);
	\draw[fill=black](2,0)circle(2pt);
	\draw[fill=black](2,1)circle(2pt);
	\draw[fill=black](2,2)circle(2pt);
	\draw[fill=black](2,3)circle(2pt);	
	\draw[fill=black](2,4)circle(2pt);	
	\draw[fill=black](3,0)circle(2pt);
	\draw[fill=black](3,1)circle(2pt);	
	\draw[fill=black](3,2)circle(2pt);
	\draw[fill=black](3,3)circle(2pt);	
	\draw[fill=black](3,4)circle(2pt);
	\draw[fill=black](4,0)circle(2pt);
	\draw[fill=black](4,1)circle(2pt);	
	\draw[fill=black](4,2)circle(2pt);
	\draw[fill=black](4,3)circle(2pt);	
	\draw[fill=black](4,4)circle(2pt);
	
	\draw[](0.1,0.1)node[cross, red]{};
	\draw[](0.3,1.9)node[cross, red]{};
	\draw[](0.1,2.5)node[cross, red]{};
	\draw[](0.9,3.1)node[cross, red]{};	
	\draw[](0.5,3.8)node[cross, red]{};	
	\draw[](1.9,0.1)node[cross, red]{};
	\draw[](1.1,1.1)node[cross, red]{};	
	\draw[](1.6,2.6)node[cross, red]{};
	\draw[](1.4,3.1)node[cross, red]{};	
	\draw[](1.7,2.9)node[cross, red]{};
	\draw[](2.9,0)node[cross, red]{};
	\draw[](2.1,1.9)node[cross, red]{};
	\draw[](2.2,2.2)node[cross, red]{};
	\draw[](2.3,3.9)node[cross, red]{};	
	\draw[](2.8,1.1)node[cross, red]{};	
	\draw[](3,0.9)node[cross, red]{};
	\draw[](3.1,1.4)node[cross, red]{};	
	\draw[](3.6,2.9)node[cross, red]{};
	\draw[](3.3,3.8)node[cross, red]{};	
	\draw[](3.1,3.1)node[cross, red]{};
	\draw[](3.1,0.2)node[cross, red]{};
	\draw[](2.8,1.3)node[cross, red]{};	
	\draw[](1.3,2.8)node[cross, red]{};
	\draw[](2.1,3.1)node[cross, red]{};	
	\draw[](4,4)node[cross, red]{};
	
	
\draw[fill=black](5.5, 3)circle(2pt) node[right]{\small $\in \mathcal{G}$};
\draw[](5.5, 1)node[cross, red]{} node[right]{\small $\in \mathcal{F}^{(c-1)}_{MC}$};

\end{tikzpicture}
\caption{\label{fig:frame_fc} The set of points in the current frame $F_c$ is given. The task is to resample the point set located at arbitrarily distributed mesh positions $\mathcal{F}^{(c)} = \mathcal{F}^{(c-1)}_{\text{MC}}$ which is denoted by the red crosses onto the regularly spaced grid positions $\mathcal{G}$ which are given as black dots.}
\end{figure}
The resampling process can be conducted using common interpolation techniques like e.g. linear or cubic interpolation. As these techniques do not promise best results, we use Key Point Agnostic Frequency-Selective Mesh-to-Grid Resampling (AFMSR) which shows large improvements in terms of quality compared to common interpolation techniques for several kinds of affine transforms \cite{Heimann_2020_MMSP}. For affine transforms, the density of the mesh points is distributed uniformely, whereas for motion compensated applications, the mesh points are not distributed uniformely but with varying density within a frame. We will evaluate the effect of this in the following.
\subsection{AFSMR}
\label{sec:afsmr}
AFSMR is a high-quality reconstruction technique. It assumes a set of horizontal  and vertical  coordinates at arbitrary non-integer positions with according color value, i.e., $\mathcal{F}^{(c)}$. These points are resampled onto regularly spaced integer grid positions $\mathcal{G}$. The point sets are further demonstrated in Figure \ref{fig:frame_fc}. AFSMR exploits the assumption that an image can be locally represented as weighted superpositions of two-dimensional basis functions $\varphi_{(k,l)}$. Hence, the image is partitioned into blocks with surrounding support area so that it holds in total $M\times N$ pixels. The image signal within this reconstruction area $\mathcal{A}$ is defined as 
\begin{equation}
\label{Eq:image}
f[m, n] = \sum_{(k, l) \in \mathcal{K}} c_{(k, l)} \varphi_{(k, l)}[m, n], 
\end{equation}
with expansion coefficients $c_{(k,l)}$, the set of available basis functions $\mathcal{K}$ and pixel coordinates $(m,n)\in \mathcal{A}$. As we use real-valued two-dimensional DCT basis functions, the expansion coefficients can be seen as transform coefficients. \\
The model is generated iteratively within the reconstruction area $\mathcal{A}$. We initialize the model $g^{(0)}$ to zero, i.e. ${g^{(0)}[m,n] = 0}$. In the $\nu-$th iteration, the model is given as
\begin{equation}
g^{(\nu)}[m, n] = g^{(\nu -1)}[m, n] + \hat{c}_{(u, v)} \varphi_{(u, v)}[m, n],
\end{equation}
where $\varphi_{(u, v)}[m, n]$ denotes the selected basis function in iteration $\nu$ and $\hat{c}_{(u, v)}$ the corresponding estimated expansion coefficient. In order to find the best fitting basis function, the residual $r^{(\nu)}$ is determined first
\begin{equation}
\label{Eq:ResidualNew}
r^{(\nu)} = f[m,n] - g^{(\nu)}[m,n].
\end{equation}
Given the residual, we can determine the residual energy
\begin{equation}
E^{(\nu)} = \sum_{(m,n) \in \mathcal{A}} w[m,n]\left(r^{(\nu)}[m,n]\right)^2,
\end{equation}
which includes the spatial weighting function $w[m,n]$. The task for the model generation process is to minimize the residual energy function as good as possible. The included spatial weighting function pays more attention to the points in the center of the reconstruction area than towards the border regions. Therefore, it is designed as an isotropically decaying window function
\begin{equation}
\label{Eq:SpatialWeightingNew}
w[m, n] = 
\rho^{\sqrt{(m-\frac{M-1}{2})^2 + (n-\frac{N-1}{2})^2}}.
\end{equation}
Now, the expansion coefficient $\hat{c}^{(\nu)}_{(k, l)}$ can be determined for all frequencies $(k, l)$. Therefore, we follow \cite{2005_Kaup_FSEOrig} and set the derivative of $E^{(\nu)}$ with respect to $\hat{c}^{(\nu)}$ to zero and get
\begin{equation}
\label{Eq:estExpCoeffNew}
\hat{c}^{(\nu)}_{(k, l)} = \frac{\sum_{(m,n)\in \mathcal{A}} r^{(\nu-1)}[m,n] \varphi_{(k, l)}[m,n]w[m,n]}{\sum_{(m,n)\in \mathcal{A}} w[m,n](\varphi_{(k, l)}[m,n])^2}.
\end{equation}
At the end of an iteration, the frequency indexes of the best fitting basis function are selected according to
\begin{equation}
{(u, v)} = \underset{{(k, l)}}{\mathrm{argmax}} \left( \Delta E_{(k, l)}^{(\nu)} w_{f}(k,l) \right),
\end{equation}
where $w_{f}(k,l)$ is the spectral weighting function. Supplemental high frequencies in the model lead to undesired artifacts like ringing and noise because natural images mainly contain low frequencies \cite{2000_Lam_DCTCoeffAnalysis}. Nevertheless, if high frequencies are dominant they still have to be included in the model generation in order to reconstruct fine structures. Therefore, the spectral weighting function is expressed as
\begin{equation}
\label{Eq:FreqWeighting}
w_{f}(k,l) = \sigma^{\sqrt{k^2 + l^2}},
\end{equation}
where $\sigma \in ]0,1[$ parametrizes the decay and $k$, $l$ denote the horizontal and vertical frequency indexes, respectively. The isotropc window function decays with increasing frequency index. Thus, it favors low frequencies more than high ones but they are still allowed to be included in the model to reconstruct fine structures. \\
After the best fitting basis function is selected, the model generation process continues with the next iteration $\nu+1$. This process is repeated until a stopping criterium like e.g. a maximum number of iterations is met. In the end, the result on the grid positions still has to be obtained. Therefore, the model offers the estimated coefficients $\hat{c}_{(k, l)}$ for all possible basis functions. These coefficients are inserted into \eqref{Eq:image}. The image signal on the grid positions $\mathcal{G}$ is then obtained by
\begin{equation}
\label{Eq:ResampledImage}
f[m, n] = \sum_{(k, l) \in \mathcal{K}} \hat{c}_{(k, l)} \varphi_{(k, l)}[m, n], \;\; \forall (m,n) \in \mathcal{G}.
\end{equation} 
Finally, the image signal $f[m, n]$ on the grid is taken while ommiting the original mesh points.

\section{Evaluation}
\label{sec:evaluation}

\begin{table*}[t]
\centering
\caption{\label{tab:res_psnr} Average results in terms of PSNR in dB. Best results are given in bold.}
\begin{tabular}{|l|c|c|c|c||c|c|c|c|}
\hline
{}			& \multicolumn{4}{c||}{Farneback}  & \multicolumn{4}{c|}{RAFT} 	\\
\hline
{} 			& 	BasketballPass	& BlowingBubble 	& 	BQ$\square$	& Horses & BasketballPass	& BlowingBubble	& 	BQ$\square$ &Horses		\\
\hline
LIN 		& 32.5 & 29.4 & 27.7 & \textbf{26.4} & 20.3 & 22.5 & 21.5 & 16.7	\\
\hline
CUB 		& 32.5 & 29.0 & 28.0 & 26.2 & 20.2 & 22.3 & 21.4 & 16.5	\\
\hline
NWE 		& 31.9 & 29.9 & 29.8 & 26.0 & 20.8 & \textbf{22.9} & \textbf{24.4} &  16.7 \\
\hline
FSMR 	& 32.0 & 29.7 & 32.2 & 25.9 & 20.7 & 22.4 & 21.9 & 16.7 \\
\hline
AFSMR 	& \textbf{33.2} & \textbf{31.9} & \textbf{35.4} & 26.3 & \textbf{20.9} & 22.8 & 22.2 & 	\textbf{16.9}\\
\hline
\end{tabular}
\end{table*}

\begin{table*}[t]
\centering
\caption{\label{tab:res_ssim} Average results in terms of SSIM. Best results are given in bold.}
\begin{tabular}{|l|c|c|c|c||c|c|c|c|}
\hline
{}			& \multicolumn{4}{c||}{Farneback}  & \multicolumn{4}{c|}{RAFT} 	\\
\hline
{} 			& 	BasketballPass	& BlowingBubble 	& 	BQ$\square$	& Horses & BasketballPass	& BlowingBubble	& 	BQ$\square$ &Horses		\\
\hline
LIN 		& 	\textbf{0.929} & 0.897 & 0.949 & \textbf{0.810} &0.618 & \textbf{0.636} & 0.779 & 	\textbf{0.400}\\
\hline
CUB 		& 	 0.928 & 0.900 & 0.954 & 0.808 & 0.611 & 0.624 & 0.775 & 0.389\\
\hline
NWE 		&  0.923 & 0.890 & 0.955 & 0.797 &  0.618 & 0.629 &	\textbf{0.806} & 0.390 \\
\hline
FSMR 	& 	0.925 & 0.884 & 0.963 & 0.798 & 0.614 & 0.612 & 0.754 & 0.390	\\
\hline
AFSMR 	& 	 0.928 & \textbf{0.913} & \textbf{0.972} & 0.807 &  \textbf{0.620} & 0.618 & 0764 & 0.391	\\
\hline
\end{tabular}
\end{table*}

We evaluate our proposed framework for the sequences of ClassD of the HEVC test set. For motion estimation, we use on the one hand the optical flow by Farneback \cite{Farneback_2003} and on the other hand one of the currently best performing motion field estimators on the Sintel data set \cite{sintel} which is neural network based and called 'RAFT' \cite{Teed_2020}. We evaluated the results for the luma channel in terms of quality using the average PSNR and SSIM for every second frame out of the first 100 of the respective sequence. Additionally, we show run times of the different interpolation techniques. 

\subsection{Quality}
\begin{figure}
\vspace{-.4cm}
\centering
\setlength\fwidth{0.41\textwidth}
%
\begin{tikzpicture}

\begin{axis}[%
width=\fwidth,
height=0.378\fwidth,
at={(0\fwidth,0\fwidth)},
scale only axis,
axis on top,
xmin=0.5,
xmax=180.5,
tick align=outside,
y dir=reverse,
ymin=0.5,
ymax=68.5,
axis line style={draw=none},
ticks=none,
legend style={legend cell align=left, align=left, draw=white!15!black}
]
\addplot [forget plot] graphics [xmin=0.5, xmax=180.5, ymin=0.5, ymax=68.5] {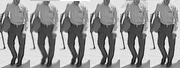};
\end{axis}

\end{tikzpicture}%
\caption{\label{fig:visualExample} An extract from frame 62 of BQSquare from left to right, original, LIN (25.9dB), CUB (25.3dB), NWE (25.0dB), FSMR (27.6dB), AFSMR (28.9dB). Best viewed enlarged.} 
\end{figure}

The evaluation results in terms of PSNR are given in Table~\ref{tab:res_psnr} and in terms of SSIM in Table~\ref{tab:res_ssim}. The results are shown for the four sequences of ClassD, namely BasketballPass, BlowingBubbles, BQSquare (BQ$\square$), and RaceHorses (Horses). For interpolation, the commonly used linear (LIN) \cite{2002_Amidror_InterpolationMethodsSurvey} and cubic (CUB) \cite{2002_Amidror_InterpolationMethodsSurvey} interpolation, as well as the Nadaraya-Watson Estimator (NWE) \cite{2007_Takeda} are analyzed. Furthermore, results are shown for the already high performing FSMR \cite{2017_Koloda_FSMR}. Last, our approach, AFSMR \cite{Heimann_2020_MMSP}, is analyzed in further detail. Moreover, the results after applying the optical Flow by Farneback are given in the left half of Table~\ref{tab:res_psnr} and Table~\ref{tab:res_ssim} and the reconstruction results after estimating the motion by RAFT is given in the right half of the tables. \\
Beginning with the Farneback motion estimation, one can see, that AFSMR performs best in nearly all cases. It can increase the high quality results from FSMR further by up to 3.2~dB for the BQSquare sequence. FSMR uses so called key points for the model generation process. These points are an estimation of the final result using cubic interpolation. Thus, the model generated using FSMR refines the results of cubic interpolation. Hence, FSMR works well in cases where cubic interpolation performs well. We do not use these key points for AFSMR. Thereby, the results are absolutely independent from the performance of other interpolation techniques. Thus, our model can generalize better and the high quality model is more stable with respect to the distribution of the mesh points. Only in the case of the RaceHorses sequence, LIN is slightly better than AFSMR. Here, the influence of inaccurate motion estimation is visible. Therefore, we used the pretrained RAFT as second motion estimator. For both, PSNR and SSIM, the qualities drop drastically. RAFT mainly focuses on the motion field estimation and thus, is inaccurate in the per pixel true motion estimation. Nevertheless, AFSMR can compete with the comparing methods. Focusing again on a good per pixel motion estimation, the results for the BQSquare sequence and Farneback motion estimation are given per frame in Figure~\ref{fig:interpolationComparison} in terms of PSNR relative to LIN. The quality of FSMR drops in some cases due to heavy ringing artifacts. For all reconstructed frames, AFSMR is the best performing method. It can increase the quality of LIN for frame 62 by nearly 10dB. An excerpt from this frame is shown in Figure~\ref{fig:visualExample}. For LIN, CUB, and NWE the walking man is blurry. FSMR and AFSMR deliver much sharper results with some ringing next to the legs with slightly less ringing artifacts for AFSMR.
\begin{figure}
\vspace{-.3cm}
\centering
\setlength\fwidth{0.38\textwidth}
%
\definecolor{mycolor1}{rgb}{0.00000,0.44700,0.74100}%
\definecolor{mycolor2}{rgb}{0.85000,0.32500,0.09800}%
\definecolor{mycolor3}{rgb}{0.92900,0.69400,0.12500}%
\definecolor{mycolor4}{rgb}{0.49400,0.18400,0.55600}%
\begin{tikzpicture}

\begin{axis}[%
width=0.951\fwidth,
height=0.59\fwidth,
xmin=0,
xmax=100,
xlabel style={at={(ticklabel cs:0.5, 0.5)}, font=\color{white!15!black}},
xlabel={\normalsize frame \#},
xtick distance = 20,
ymin=-15,
ymax=10,
ylabel style={at={(ticklabel cs:0.5, 0.5)}, font=\color{white!15!black}},
ylabel={\normalsize $\Delta$ PSNR in dB},
ytick distance = 10,
axis background/.style={fill=white},
axis x line*=bottom,
axis y line*=left,
legend style={at={(0,0)}, anchor=south west, legend columns=4, legend cell align=left, align=left, draw=white!15!black}
]
\addplot [line width = 0.5mm, color=mycolor1]
  table[row sep=crcr]{%
2	-0.0386160030004952\\
4	-0.552586126169942\\
6	0.126856795121991\\
8	-1.72051892446051\\
10	-0.42334364371191\\
12	0.246890293147089\\
14	0.297752774456352\\
16	0.133535497745729\\
18	0.139915714595233\\
20	0.117673884561565\\
22	0.215219080636601\\
24	-0.0644628411706307\\
26	-0.03463961188816\\
28	-0.0664752397522896\\
30	-0.22279345503733\\
32	-0.102755057608448\\
34	-0.121020428037657\\
36	-0.0686059574056301\\
38	-0.111337594891022\\
40	-0.063538661262978\\
42	-0.0870737664362835\\
44	-0.0909680741231931\\
46	-0.0865715255632402\\
48	-0.32631713598207\\
50	-0.340874238470178\\
52	-0.545655288189835\\
54	-1.55319082501917\\
56	-1.57930547110426\\
58	-0.21935586544263\\
60	-0.51735965345603\\
62	-0.911425403502435\\
64	-1.15824090288796\\
66	-0.507005757689129\\
68	0.0411707941686608\\
70	0.365992237716977\\
72	-0.885819529935894\\
74	-2.09092552373492\\
76	-0.0383809646082902\\
78	-0.430070850984354\\
80	-0.515570251590688\\
82	-0.987134527340089\\
84	-0.873720387758972\\
86	-1.53168952213756\\
88	-1.2222453865538\\
90	-0.935190434599676\\
92	-0.0867282820550912\\
94	-0.768832328006113\\
96	-0.0541240654717896\\
98	-0.836669313994467\\
100	0.0186927446179581\\
};
\addlegendentry{CUB}

\addplot [line width = 0.5mm, color=mycolor2]
  table[row sep=crcr]{%
2	-0.291578079900383\\
4	-0.281657967772368\\
6	0.269176755602192\\
8	-0.760037438478456\\
10	-1.00066027052186\\
12	-1.21957823908587\\
14	-0.421503413321958\\
16	-0.706555278008139\\
18	0.266168189247018\\
20	0.483608733739846\\
22	0.362127036843674\\
24	-0.546199148859401\\
26	-0.299407074609015\\
28	-0.259522576209772\\
30	-0.00514668360560577\\
32	-0.130248647787411\\
34	-0.158941852620188\\
36	-0.0980663734644196\\
38	-0.0536791461004391\\
40	0.565430360041017\\
42	-0.141330952056368\\
44	0.524794499328621\\
46	1.0206442489817\\
48	-0.034241669534719\\
50	0.966549527945205\\
52	1.708463375877\\
54	-0.445913001235667\\
56	0.580404496656371\\
58	1.33940378145032\\
60	4.38963235095274\\
62	5.95781054956075\\
64	0.495607871796004\\
66	1.26519797154077\\
68	1.47450530632165\\
70	1.65141409218102\\
72	-0.171172039509656\\
74	0.498539158907128\\
76	4.19968300682057\\
78	0.415393172625663\\
80	0.856223288911679\\
82	-0.456857502909031\\
84	0.0127955492891729\\
86	-0.723868887961871\\
88	-0.303207833746409\\
90	-0.74131600277456\\
92	-0.112929451311004\\
94	1.08839836381189\\
96	0.45235277318308\\
98	2.21254305329762\\
100	2.53057093448344\\
};
\addlegendentry{NWE}

\addplot [line width = 0.5mm, color=mycolor3]
  table[row sep=crcr]{%
2	0.0368236539903393\\
4	-0.107463040205069\\
6	1.01790305087745\\
8	-0.262034809126206\\
10	-0.778774193062119\\
12	0.799076947262641\\
14	1.23414412472482\\
16	0.731874607306654\\
18	1.36010600339323\\
20	1.02066043502628\\
22	1.18400699707041\\
24	-11.2566400684255\\
26	-6.77414674008453\\
28	-0.193663297980322\\
30	-0.598773059825394\\
32	-0.210118168064682\\
34	-0.198663494087544\\
36	-0.179711877478205\\
38	-0.138921022656035\\
40	0.428408114281993\\
42	-0.154398523646417\\
44	0.470251755228023\\
46	0.871086623804779\\
48	-0.22876036304876\\
50	0.476719837904994\\
52	1.59975470334411\\
54	-1.55393134993543\\
56	-7.30538497764346\\
58	2.17605163613249\\
60	4.56683223439091\\
62	-2.20256913177276\\
64	-2.35954646367291\\
66	2.77729219361268\\
68	2.80939460092437\\
70	2.62041261486014\\
72	1.05736083198748\\
74	0.0791297540919018\\
76	4.86895352744808\\
78	1.67981602323416\\
80	2.24609642585857\\
82	1.24348745135357\\
84	1.45536885051484\\
86	0.196414339992948\\
88	0.33262212580361\\
90	0.712017151936763\\
92	0.613493915548432\\
94	2.2623142794444\\
96	2.31255114576731\\
98	1.63764048179428\\
100	3.47232552711807\\
};
\addlegendentry{FSMR}

\addplot [line width = 0.5mm, color=mycolor4]
  table[row sep=crcr]{%
2	0.934013201209922\\
4	1.42380802416595\\
6	1.68703228013766\\
8	1.21162204427706\\
10	1.68302521002074\\
12	1.84373397824001\\
14	2.20627527881846\\
16	1.69119830659126\\
18	2.26132926367886\\
20	2.05422797679534\\
22	2.45441982467162\\
24	0.691286065811209\\
26	0.0460024904793919\\
28	-0.0256302410222844\\
30	0.198060572919928\\
32	-0.0325342140840874\\
34	-0.0586713776956032\\
36	-0.0375244739171663\\
38	-0.055909407133818\\
40	0.601350470859771\\
42	-0.0723312229422568\\
44	0.601086950353601\\
46	1.18553019006441\\
48	0.279762498167614\\
50	1.30288299598593\\
52	4.59986562438888\\
54	2.46450222507729\\
56	3.64820665979376\\
58	4.46334825251763\\
60	7.12400479099867\\
62	9.26904797111331\\
64	4.88004622823502\\
66	5.56879550137105\\
68	5.32160490102715\\
70	4.94306675284254\\
72	3.28704353658217\\
74	2.32147779888198\\
76	6.66048879729527\\
78	3.80309738090715\\
80	4.3746402041138\\
82	3.01512600340226\\
84	3.2415596273889\\
86	2.83873900484348\\
88	1.3450037559496\\
90	2.0439581172577\\
92	2.41027408549992\\
94	4.16389193925329\\
96	3.76303943048833\\
98	3.84732824718471\\
100	4.36774479252283\\
};
\addlegendentry{AFSMR}

\end{axis}
\end{tikzpicture}%
\caption{\label{fig:interpolationComparison} Quality of BQSquare sequence using Farneback motion estimation for all reconstructed images relative to LIN.\vspace{-.2cm}} 
\end{figure}

\subsection{Run Time}
As a second evaluation, we show the run times for the various interpolation techniques. Therefore, we average the run times of the Basketball Pass sequence from ClassD. The results are given in Table \ref{tab:res_time}. All evaluations are pursued on a mid-range computer equipped with an \textit{Intel Xeon(R) CPU E3-1275} with 3.8GHz. Obviously, the commonly used interpolation techniques linear and cubic interpolation are the fastest techniques. The high quality model-based approaches, FSMR and AFSMR, cannot compete with them as non-optimized MATLAB implementations are used. Nevertheless, AFSMR improves the run time by 11 times relative to FSMR. Hence, it improves FSMR in terms of run time as well as in terms of quality at the same time for FRUC applications.
\begin{table}
\centering
\vspace{-.3cm}
\caption{\label{tab:res_time} Average run times for BasketballPass in sec. Speed-up factor given relative to FSMR. }
\begin{tabular}{|l|c|c|}
\hline
{} & Run time & Speed-up factor \\
\hline
LIN 			& 0.7 & 2253 \\
\hline
CUB 			& 0.8 & 1971 \\
\hline
NWE 			& 15.6 & 101 \\
\hline
FSMR 		& 1577.1 & ---\\
\hline
AFSMR 		& 140.9 & 11 \\
\hline
\end{tabular}
\vspace{-.3cm}
\end{table}
\vspace{-.2cm}
\section{Conclusion}
\label{sec:conclusion}
\vspace{-.2cm}
Frame Rate Up-Conversion is of high interest in many application areas. Using a unidirecitonal motion compensated approach, pixels from the previous frame are shifted into the to be reconstructed frame onto arbitrary non-integer positions. Thus, a mesh with varying pixel distribution emerges. Hence, a resampling from mesh to regularly spaced grid positions is necessary to display the result on a digital screen. Common interpolation like linear or cubic interpolation can be used but they do not show the best results in terms of quality for a good true motion estimation. Therefore, a more sophisticated approach has to be used. We are using AFSMR that improves the high-quality results from FSMR for up to 3.2~dB in terms of PSNR and increases the run time by 11 times simultaneously.
\vspace{-.2cm}
\section{Acknowledgment}
The authors gratefully acknowledge that this work has been supported by the Deutsche Forschungsgemeinschaft (DFG) under contract number KA 926/8-1.

\vfill\pagebreak

\bibliographystyle{IEEEbib}
\bibliography{bib_4icassp2021}

\begin{thebibliography}{10}

\bibitem{Ohm_2016}
Jens-Rainer Ohm,
\newblock {\em Multimedia Content Analysis},
\newblock Springer-Verlag Berlin Heidelberg, 2016.

\bibitem{Bao_2019}
Wenbo Bao, Wei-Sheng Lai, Chao Ma, Xiaoyun Uhang, Zhiyong Gao, and Ming-Hsuan
  Yang,
\newblock ``{Depth-Aware Video Frame Interpolation},''
\newblock in {\em Proceedings of the IEEE Conference on Computer Vision and
  Pattern Recognition}, 2019.

\bibitem{Zhang_2020}
Y.~{Zhang}, L.~{Chen}, C.~{Yan}, P.~{Qin}, X.~{Ji}, and Q.~{Dai},
\newblock ``{Weighted Convolutional Motion-Compensated Frame Rate Up-Conversion
  Using Deep Residual Network},''
\newblock {\em IEEE Transactions on Circuits and Systems for Video Technology},
  vol. 30, no. 1, pp. 11--22, 2020.

\bibitem{Haan_2010}
G.~de~Haan,
\newblock {\em {Digital Video Post Processing}},
\newblock 2010.

\bibitem{Batz_2017}
Michel B\"atz, Fabian Brand, Andrea Eichenseer, and Andr\'e Kaup,
\newblock ``{Motion Compoensated Frame Rate Up-Conversion using 3D Frequency
  Selective Extrapolation and a Multi-Layer Consistensy Check},''
\newblock in {\em Proceedings of the IEEE International Conference on
  Acoustics, Speech and Signal Processing}, 2017.

\bibitem{Kaviani_2016}
H.~R. {Kaviani} and S.~{Shirani},
\newblock ``{Frame Rate Upconversion Using Optical Flow and Patch-Based
  Reconstruction},''
\newblock {\em IEEE Transactions on Circuits and Systems for Video Technology},
  vol. 26, no. 9, pp. 1581--1594, 2016.

\bibitem{Teed_2020}
Zachary Teed and Jia Deng,
\newblock ``{RAFT: Recurrent All-Pairs Field Transforms for Optical Flow},''
\newblock in {\em Proceeding of the 16th European Conference on Computer
  Vision}, 2020.

\bibitem{Heimann_2020_MMSP}
Viktoria Heimann, Nils Genser, and Andr\'e Kaup,
\newblock ``{Key Point Agnostic Frequency-Selective Mesh-to-Grid Resampling
  using Spectral Weighting},''
\newblock in {\em Proceedings of the IEEE 22nd Workshop on Multimedia Signal
  Processing}, 2020.

\bibitem{2005_Kaup_FSEOrig}
Andr\'e Kaup, K.~Meisinger, and T.~Aach,
\newblock ``{Frequency selective signal extrapolation with applications to
  error concealment in image communication},''
\newblock in {\em International Journal of Electronics and Communications},
  June 2005, vol.~59, pp. 147--156.

\bibitem{2000_Lam_DCTCoeffAnalysis}
Edmund~Y. Lam and Joseph~W. Goodman,
\newblock ``{A Mathematical Analysis of the DCT Coefficient Distributions for
  Images},''
\newblock in {\em IEEE Transactions on Image Processing}, October 2000, vol.~9,
  pp. 1661--1666.

\bibitem{Farneback_2003}
{G. Farneback},
\newblock ``{Two-Frame Motion Estimation Based on Polynomial Expansion},''
\newblock in {\em Proceedings of the 13th Scandinavian Conference on Image
  Analysis}, 2003, pp. 363--370.

\bibitem{sintel}
Sintel,
\newblock ``Results and rankings,''
  \url{http://sintel.is.tue.mpg.de/quant?metric_id=0&selected_pass=0}.

\bibitem{2002_Amidror_InterpolationMethodsSurvey}
Isaac Amidror,
\newblock ``{Scattered data interpolation methods for electronic imaging
  systems: A survey},''
\newblock in {\em Journal of Electronic Imaging}, April 2002, vol.~11, pp.
  157--176.

\bibitem{2007_Takeda}
Hiroyuki Takeda, S.~Farsiu, and P.~Milanfar,
\newblock ``Kernel regression for image processing and reconstruction,''
\newblock in {\em IEEE Transactions on Image Processing}, February 2007,
  vol.~16, pp. 349--366.

\bibitem{2017_Koloda_FSMR}
J\'an Koloda, J.~Seiler, and A.~Kaup,
\newblock ``{Frequency-Selective Mesh-to-Grid Resampling for Image
  Communication},''
\newblock in {\em IEEE Transactions on Multimedia}, August 2017, vol.~19, pp.
  1689--1701.

\end{thebibliography}

\end{document}